%% file: ratiopaper_techreport_arxiv.tex
\documentclass[11pt, onecolumn, a4paper, dvips]{article}

\usepackage[margin=3cm]{geometry}
\usepackage{times}
\usepackage[english]{babel}
\usepackage{amsmath, amsfonts, bm, amssymb, amsthm}
\usepackage{graphicx,psfrag,color}
\graphicspath{{figures/}}
\usepackage[numbers, sort&compress]{natbib}
\usepackage{enumerate}

\usepackage{color}
\definecolor{linkcol}{rgb}{0,0,0.4} 
\definecolor{citecol}{rgb}{0.5,0,0}
\usepackage{hyperref}
\hypersetup
{
  bookmarksopen=true,
  breaklinks=true,
  hyperindex=true,
  pdfhighlight=/O,   
  colorlinks=true,   
  linkcolor=linkcol, 
  citecolor=citecol, 
  urlcolor=linkcol   
}
\hypersetup
{
  pdftitle={Note on the computation of the Metropolis-Hastings 
    ratio for Birth-or-Death moves in trans-dimensional MCMC 
    algorithms for signal decomposition problems},
  pdfauthor={Alireza Roodaki, Julien Bect and Gilles Fleury}, 
  pdfsubject={},
  pdfkeywords={MCMC, RJ-MCMC, ratio, birth-or-death, signal
    decomposition, trans-dimensional}
}

\newtheorem{prop}{Proposition}

\newenvironment{remark}[1][Remark]{\begin{trivlist}
\item[\hskip \labelsep {\bfseries #1}]}{\end{trivlist}}

\DeclareMathAlphabet\PazoBB{U}{fplmbb}{m}{n}

\newcommand \thk {\bm{\theta}_k}
\newcommand \y   {\vect{y}}

\newcommand \ak   {\vect{a}_k}
\newcommand \ds   {\delta^2}

\newcommand \ok   {\bm{\omega}_k}
\newcommand \D    {\vect{D}}
\newcommand \DtD  {\vect{D}_k^t \vect{D}_k}

\newcommand \yPy  [1]{\vect{y}^t\vect{P}_{#1}\vect{y}}
\newcommand \Xset   {\mathbb{X}}
\newcommand \Sset   {\mathbb{S}}
\newcommand \Bor  {\mathcal{B}}
\newcommand \WW   {\mathbb{W}}
\newcommand \MM   {\mathbb{M}}
\newcommand \Nset {\mathbb{N}}
\newcommand \Rset {\mathbb{R}}

\newcommand \tth {\bm{\theta}}
\renewcommand \ss {\bm{s}}
\newcommand \xx {\bm{x}}

\newcommand \bb {\mathrm{b}}
\newcommand \dd {\mathrm{d}}

\newcommand \ddiff   {\mathrm{d}}
\newcommand \dxx     {\ddiff\xx}
\newcommand \diffs   {\ddiff s}
\newcommand \diffss  {\ddiff \ss}

\newcommand \vect[1]  {\mathbf{#1}}
\newcommand \un       {\PazoBB{1}}
\newcommand \model    {\mathcal{M}}
\newcommand \kmax     {k_{\text{max}}}
\newcommand \Kcal     {\mathcal{K}}

\newcommand \tr {{\mathrm{T}}}
\newcommand \dotvar {\,\bm{\cdot}\,}

\title{\textsc{Technical report}\\[1.5em]
  Note on the computation of the Metropolis-Hastings ratio\\
  for Birth-or-Death moves in trans-dimensional MCMC\\
  algorithms for signal decomposition problems}

\author{%
  Alireza~Roodaki%
  \thanks{
    Corresponding authors.%
  }
  $\mskip 1mu{}^, \mskip -5mu$
  \thanks{
    The results presented here are also part of the PhD thesis of the
    first author \cite[Chapter 1]{roodaki-2012-phd}.%
  } 
  $\mskip 1mu{}^, \mskip -5mu$  
  \thanks{
    Email: alireza4702@gmail.com.%
  }
  \and%
  Julien~Bect%
  \footnotemark[1]
  $\mskip 1mu{}^, \mskip -5mu$  
  \thanks{
    Email: \{julien.bect, gilles.fleury\}@supelec.fr%
  }
  \and%
  Gilles~Fleury
  \footnotemark[4] 
}

\date{%
  E3S --- \textsc{Supelec Systems Sciences}\\
  Department of Signal Processing and Electronic Systems\\
  \textsc{SUPELEC}, Gif-sur-Yvette, France.
}

\begin{document}
\maketitle


\begin{abstract}
  Reversible jump MCMC (RJ-MCMC) sampling techniques, which allow to
  jointly tackle model selection and parameter estimation problems in
  a coherent Bayesian framework, have become increasingly popular in
  the signal processing literature since the seminal paper of Andrieu
  and Doucet (\emph{IEEE Trans. Signal Process.}, 47(10),
  1999). Crucial to the implementation of any RJ-MCMC sampler is the
  computation of the so-called Metropolis-Hastings-Green (MHG) ratio,
  which determines the acceptance probability for the proposed moves.

  It turns out that the expression of the MHG ratio that was given in
  the paper of Andrieu and Doucet for ``Birth-or-Death'' moves---the
  simplest kind of trans-dimensional move, used in virtually all
  applications of RJ-MCMC to signal decomposition problems---was
  erroneous. Unfortunately, this mistake has been reproduced in many
  subsequent papers dealing with RJ-MCMC sampling in the signal
  processing literature.

  This note discusses the computation of the MHG ratio, with a focus
  on the case where the proposal kernel can be decomposed as a mixture
  of simpler kernels, for which the MHG ratio is easy to compute. We
  provide sufficient conditions under which the MHG ratio of the
  mixture can be deduced from the MHG ratios of the elementary kernels
  of which it is composed. As an application, we consider the case of
  Birth-or-Death moves, and provide a corrected expression for the
  erroneous ratio in the paper of Andrieu and Doucet.

\end{abstract}

\section{Introduction}
\label{sec:introduction}

Model selection and parameter estimation are fundamental tasks arising
in many (if not all) signal processing problems, when parametric
models are employed. Let us consider a collection of models $\{
\model_k,\, k\in \Kcal \}$, indexed by some finite or countable
set~$\Kcal \subset \Nset$, with parameter vector $\thk \in \Theta_k \subset
\Rset^{n_k}$ under model~$\model_k$. In a Bayesian framework, model
selection (or averaging) and parameter estimation can in principle be
carried out jointly, using the posterior distribution of the
pair~$\left( k, \thk \right)$,
\begin{equation}
  \pi\left(k,\,\thk \right) %
  \;\propto\; %
  p\left(\y\,|\,k,\,\thk\right)\, p\left(k,\,\thk\right),
\end{equation}
where $\y$ is the observed data and $\propto$ indicates
proportionality. Note that the distribution~$\pi$ is defined on $\Xset
= \bigcup_{k\in\Kcal}\, \{k\} \times \Theta_k$, which is a disjoint
union of subspaces with differing dimensionality. Generic Markov Chain Monte
Carlo (MCMC) methods for probability distributions defined on such
spaces became available during the 90's, most notably Green's widely
applicable RJ-MCMC sampler \cite{green:1995:reversible}, making it
possible to use a fully Bayesian approach for model selection (or
averaging) and parameter estimation in all sorts of applications. The
reader is referred to \cite{andrieu:2001:msm, green:2003:trans, cappe:2003:reversible,
  sisson:2005:trans-dimensional, robert:2004:monte}
for a broader view on trans-dimensional sampling techniques (including
alternatives to the RJ-MCMC sampler).

Green's RJ-MCMC sampler can be seen as a generalization of the
well-known Metropolis-Hastings sampler~\cite{metropolis:1953:equation,
  hastings:1970:monte}, which is capable of exploring not only the
fixed-dimensional parameter spaces~$\Theta_k$, but also the
space~$\Kcal$ of all models under consideration. This algorithm relies
on an accept/reject mechanism, with an acceptance ratio
calibrated in such a way that the invariant distribution of the chain
is the target distribution~$\pi$. The computation of this acceptance
ratio for trans-dimensional moves is in general a delicate issue%
\footnote{Fortunately, the simple and powerful ``dimension
  matching'' argument \cite{green:1995:reversible} allows to bypass
  this difficulty for a large class of proposal distributions.},
involving measure theoretic considerations.

Andrieu and Doucet~\cite{andrieu:1999:jbm} pioneered the use of
RJ-MCMC sampling in ``signal decomposition'' problems, by tackling
joint model selection and parameter estimation for an unknown number
of sinusoidal signals observed in white Gaussian noise. (At the same
period, RJ-MCMC also became popular for image processing tasks such as
segmentation and object recognition; see, e.g., \cite{hurn:1997:high,
  pievatolo:1998:object, nicholls:1998:bayesian, rue:1999:bayesian,
  descombes:2001:parameter}.) This seminal papers was followed by many
others in the signal processing literature
\cite{andrieu:2000:RJMCMC-SA, andrieu:2001:robust,
  andrieu:2002:bayesian, larocque:2002:reversiblejump,
  larocque:2002:particle, ng:2005:wideband, davy:2006:bap,
  shi:2007:study, melie:2008:bayesian, ng:2008:particle,
  hong:2010:joint, schmidt:2010:inmf, rubtsov:2007:time}, relying
systematically on the original paper~\cite{andrieu:1999:jbm} for the
computation of the acceptance ratio of ``Birth-or-Death'' moves---the
most elementary type of trans-dimensional move, which either adds or
removes a component from the signal decomposition. Unfortunately, the
expression of the acceptance ratio for Birth-or-Death moves provided
by~\cite[Equation (20)]{andrieu:1999:jbm} turns out to be erroneous,
as will be explained later. Worse, the exact same mistake has been
reproduced in most of the following papers, referred to above.

The aim of this note is to provide clear statements of some
mathematical results, perhaps not completely new but never stated
explicitly, which can be used for a clean justification of the
acceptance ratio of Birth-or-Death moves in signal decomposition (and
similar) problems. Section~\ref{sec:trans} recalls, very quickly, the
basics of MCMC methods, with a focus on Metropolis-Hastings algorithms
on general state spaces (also known as RJ-MCMC
algorithms). Section~\ref{sec:mixture} discusses the computation of
the acceptance ratio for mixture kernels, and provides conditions
under which the ratio of the mixture can be directly derived from the
ratio of the elementary kernels of which it is
composed. Section~\ref{sec:bod-exchang} defines Birth-or-Death moves
and provides the expression of the ratio; several distinct but related
mathematical representations---``unsorted vectors'', ``sorted
vectors'' and Point processes---are discussed. As an illustration,
Section~\ref{sec:sinusoid} returns to the problem considered
in~\cite{andrieu:1999:jbm} and provides a corrected expression for the
Birth-or-Death ratio. Section~\ref{sec:conclusion} concludes the
paper.

\section{Background on MCMC methods}
\label{sec:trans}

This section recalls basic definitions and results for the MCMC
method. The reader is referred to~\cite{meyn:1993:markov,
  tierney:1994:markov, green:1995:reversible, gilks:1996:mcmc,
  tierney:1998:note, green:2003:trans, robert:2004:monte,
  roberts:2004:MCMC} for more detailed explanations.

\subsection{MCMC with reversible kernels}
\label{subsec:preliminary}

Let~$\pi$ be a probability distribution on a measurable space $\left(
  \Xset,\, \Bor \right)$, which is to be sampled from. MCMC sampling
methods proceed by constructing a time-homogeneous Markov
chain~$\left( \xx_n \right)$ with invariant distribution~$\pi$, using
a transition kernel $P$ that is \emph{reversible} with respect
to~$\pi$, i.e., a kernel that satisfies the detailed balance condition
\begin{equation}
  \label{eq:detailed-balance}
  \pi\left( \dxx \right) P\left( \xx, \dxx' \right)
  \;=\; \pi\left( \dxx' \right) P\left( \xx', \dxx \right).
\end{equation}
For all measurable sets $A \in \Bor$,
integrating~\eqref{eq:detailed-balance} on $\Xset \times A$ yields
\begin{equation*}
  \int_{\Xset} \pi\left( \dxx \right)\, 
  P\left( \xx,\, A \right)
  \;=\; \pi\left(A\right),
\end{equation*}
which means that $\pi$ is an invariant distribution for the kernel~$P$
(it is also said that ``$P$ leaves $\pi$ invariant'').  

If the transition kernel~$P$ is $\pi$-irreducible and aperiodic, then
\cite[Theorem~1]{tierney:1994:markov} $\pi$~is the unique invariant
distribution and the chain converges in total variation to~$\pi$ for
$\pi$-almost all starting states~$\xx$. If $P$ is also Harris
recurrent, then convergence occurs for all initial distributions
\cite[Theorem~6.51]{robert:2004:monte}.

\begin{remark}
  Some of the above requirements on the chain~$\left( \xx_n \right)$ can
  be relaxed. Most notably, time-inhomogeneous chains are used in the
  context of ``adaptive MCMC'' algorithms; see, e.g.,
  \cite{andrieu:2006,atchade:2005} and the references therein. It is
  also possible to depart from the reversibility assumption, which is a
  sufficient but not necessary condition for~$\pi$ to be an invariant
  distribution (see, e.g., \cite{diaconis:2000:nonreversible}), though
  the vast majority of MCMC algorithms considered in the literature
  are based on reversible kernels.
\end{remark}

\subsection{Metropolis-Hastings-Green kernels}
\label{sec:mhg-kernels}

The very popular Metropolis-Hastings-Green kernels, sometimes simply
called Metropolis-Hastings kernels, correspond to the following
two-stage sampling procedure: first, given that the current state of
the Markov chain is $\xx \in \Xset$, a new state $\xx' \in \Xset$ is
proposed from a transition kernel $Q\left( \xx, \,\dxx' \right)$;
second, this move is accepted with probability $\alpha\left( \xx,\,
  \xx' \right)$ and rejected otherwise---in which case the new state
is equal to~$\xx$. More formally, for all $\xx \in \Xset$ and $B \in
\Bor$, the transition kernel is given by
\begin{equation}
  \label{eq:transition}
  P\left( \xx,\, B \right) 
  \;=\; 
  \int_{B}Q\left( \xx,\,\dxx' \right)\,
  \alpha\left(\xx,\xx'\right) \;+\; 
  s\left( \xx \right) \un_B \left( \xx \right),
\end{equation}
where $\un_B$ denotes the indicator function of~$B$, and
\begin{equation*}
  s\left( \xx \right)
  \;=\;
  \int_{\Xset} Q\left( \xx,\, \dxx' \right)\,
  \left( 1 - \alpha\left(\xx,\xx' \right) \right)
\end{equation*}
is the probability of rejection at~$\xx$. It is easily seen that the
detailed balance condition~\eqref{eq:detailed-balance} holds if and
only if \cite{green:1995:reversible, tierney:1994:markov,
  tierney:1998:note}
\begin{equation}
  \label{eq:cond-alpha}
  \pi\left( \dxx \right) 
  Q\left( \xx, \dxx' \right)
  \alpha\left( \xx, \xx' \right)
  \;=\; 
  \pi\left( \dxx' \right) 
  Q\left(\xx', \dxx \right)
  \alpha\left( \xx', \xx \right).
\end{equation} 
This is achieved, for instance, by the acceptance probability
\begin{equation}
  \label{eq:acc-prob}
  \alpha\left( \xx, \xx' \right)
  \;=\;
  \min\left\{ 
    1,\, r\left(\xx,\xx'\right)
  \right\},
\end{equation}
where~$r( \xx, \xx' )$ denotes the Metropolis-Hastings-Green (MHG)
ratio
\begin{equation}
  \label{eq:acc-ratio}
  r\left( \xx, \xx' \right) 
  \;=\;
  \frac{ 
    \pi\left( \dxx' \right) Q\left( \xx', \dxx \right) 
  }{
    \pi\left( \dxx \right) Q\left( \xx, \dxx' \right)
  }\cdot
\end{equation}
The right-hand side of \eqref{eq:acc-ratio} is the Radon-Nykodim
derivative of~$\pi\left( \dxx' \right) Q\left( \xx', \dxx \right)$
with respect to~$\pi\left( \dxx \right) Q\left( \xx, \dxx' \right)$; see
\cite[Section~2]{tierney:1998:note} for technical details.

\begin{remark}
  It is proved in \cite[Section~4]{tierney:1998:note} that the
  acceptance probability~(\ref{eq:acc-prob}) is optimal in the sense
  of minimizing the asymptotic variance of sample path averages among
  all acceptance rates satisfying~(\ref{eq:cond-alpha}).
\end{remark}

\section{Mixture of proposal kernels}
\label{sec:mixture}

\subsection{Metropolis-Hastings-Green ratio for mixture of proposal
  kernels} \label{sec:ratio-mixture}

It is often convenient to consider a proposal kernel~$Q$ built as a
mixture of simpler transition kernels~$Q_m$, with $m$ in some finite
or countable index set~$\MM$. In this case we have
\begin{equation}
  \label{equ:mixt-repr}
  Q\left( \xx,\, \dxx' \right) \,=\,
  \sum_{m\in\MM}\, j\left(\xx,\,m\right)\, Q_m\left(\xx,\,\dxx'\right),
\end{equation}
where $j\left( \xx,\, m \right)$ is the probability of choosing the
move type~$m$ given that the current state is~$\xx$.
Note that the actual value of $Q_m( \xx, \dotvar )$ is irrelevant when
$j(\xx,m) = 0$.

It turns out that, under some assumptions, the MHG ratio for a mixture
kernel~$Q$ can be conveniently deduced from the elementary ratios
computed for each individual kernel~$Q_m$ using the formula
\begin{equation}
    \label{eq:acc-prob-mix}
    r\left( \xx, \xx' \right) \;=\;
    \frac{ j\left( \xx',\, m' \right) }{ j\left( \xx,\, m \right)} 
    \;
    \frac{ 
      \pi\left( \dxx' \right) Q_{m'}\left( \xx', \dxx \right)
    }{%
      \pi\left( \dxx \right) Q_m\left( \xx, \dxx'\right)
    },
\end{equation}
where $m \in \MM$ denotes the specific move that has been used to
propose~$\xx'$, and $m' \in \MM$ is the corresponding ``reverse
move''.  Equation~\eqref{eq:acc-prob-mix} is routinely used in
applications of the RJ-MCMC algorithm, and is alluded to in Green's
paper \cite[p.~717]{green:1995:reversible} in the sentence :
``\emph{If [other] discrete variables are generated in making
  proposals, the probability functions of their realized values are
  multiplied into the move probabilities}''---but it is wrong
in general. Sufficient conditions for Equation~\eqref{eq:acc-prob-mix}
to hold are provided by the following result:

\medbreak

\begin{prop}
  \label{prop:ratio}
  Let $$R_m( \dxx, \dxx' ) = j(\xx,m)\, \pi(\dxx)\, Q_m( \xx, \dxx'
  ).$$ Assume that there exists a family of disjoint sets $\WW_m \in
  \Bor \otimes \Bor$ indexed by~$\MM$ such that :
  \begin{enumerate}[i) ]
  \item For each $m \in \MM$, $R_m$ is supported by~$\WW_m$,
    which means $R_m\left( \Xset^2 \setminus \WW_m \right) = 0$.
  \item Each move~$m \in \MM$ has a unique ``reverse move''
    $\varphi(m) \in \MM$ in the sense that $\WW_{\varphi(m)} =
    \WW_m^\tr$, where $ \WW_m^\tr = \{ (\xx',\xx) : (\xx,\xx') \in
    \WW_m \}$. 
  \end{enumerate}
  Then, then MHG ratio is given by
  Equation~\eqref{eq:acc-prob-mix} with $m' = \varphi(m)$.
\end{prop}

\medbreak

\begin{proof}
  For $\pi(\dxx)\, Q(\xx,\dxx')$-almost everywhere on~$\Xset^2$, there
  is a unique $m = m_{\xx,\xx'} \in \MM$ such that $(\xx,\xx') \in
  \WW_m$. Equation~\eqref{eq:acc-prob-mix} can be rewritten as:
  \begin{equation*}
    r\left(\xx,\xx'\right) \;=\;
    \frac{ R_{\varphi(m_{\xx,\xx'})}(\dxx',\dxx) }{ R_{m_{\xx,\xx'}}(\dxx,\dxx') }\cdot
  \end{equation*}
  Then, for all $A \in \Bor \otimes \Bor$,
  \begin{align*}
    \iint_A r(\xx,&\xx')\, R(\dxx,\dxx') \\
    & =\; \iint_A \frac{ R_{\varphi(m_{\xx,\xx'})}(\dxx',\dxx) }{
      R_{m_{\xx,\xx'}}(\dxx,\dxx') }
    \,\cdot\, \sum_{m_0 \in \MM} R_{m_0}(\dxx,\dxx') \\
    & =\; \sum_{m_0 \in \MM} \iint_{A \cap \WW_{m_0}} \frac{ R_{\varphi(m_0)}(\dxx',\dxx)
    }{ R_{m_0}(\dxx,\dxx') }\, R_{m_0}(\dxx,\dxx') \\
    & =\; \sum_{m_0 \in \MM} \iint_{A\cap \WW_{m_0}}
    R_{\varphi(m_0)}(\dxx',\dxx) \\
    & =\; \sum_{m_0 \in \MM} \iint_{A^\tr \cap \WW_{m_0}^\tr}
    R_{\varphi(m_0)}(\dxx,\dxx') \\
    & =\; \iint_{A^\tr} R(\dxx,\dxx')
    \qquad \text{because } \WW_{m_0}^\tr = \WW_{\varphi(m_0)} \\
    & =\; \iint_{A} R(\dxx',\dxx) \,.
  \end{align*}
\end{proof}

\subsection{Mixture representation of trans-dimensional
  kernels} \label{sec:canonical-transdim}

Consider the case of a variable-dimensional space, that can be written
as $\Xset = \cup_{k \in \Kcal}\, \{k\} \times \Theta_k$, with $\Kcal$ a
finite or countable set (usually $\Kcal \subset \Nset$) and $\Theta_k
\subset \Rset^{n_k}$.  A point $\xx \in \Xset$ is a pair $\left( k, \tth
\right)$ with $k \in \Kcal$ and $\tth \in \Theta_k$. The problem of
sampling a (posterior) distribution on such a space typically occurs
in the context of Bayesian model selection or averaging.

Set $\Xset_k = \{ k \} \times \Theta_k$.  Any kernel~$Q$ on~$\Xset$ admits
a natural representation as a mixture of fixed-dimensional and
trans-dimensional kernels :
\begin{equation}
  \label{equ:transdim-mixture}
  Q\left( \xx, \dxx' \right) \;=\;
  \sum_{(k,l) \in \Kcal^2} p_{k,l}(\xx)\, 
  Q_{k,l}\left( \xx,\dxx' \right) \,,
\end{equation}
where
\begin{align*}
      p_{k,l}(\xx) &\;=\; \un_{\Xset_k}(\xx)\, Q( \xx,\Xset_l)\,, \\
      Q_{k,l}( \xx,\dotvar) & \;=\; \frac{1}{p_{k,l}(\xx)}\, Q\left( \xx, \dotvar
        \cap \Xset_l \right) \,.
\end{align*}
(An arbitrary value can be chosen for $Q_{k,l}( \xx,\dotvar)$ when
$p_{k,l}(\xx) = 0$ to make it a completely defined transition kernel.)
The kernels $Q_{k,k}$, $k \in \Kcal$, correspond to the
``fixed-dimensional'' part of the transition kernel~$Q$; while the
kernels $Q_{k,l}$, $(k,l) \in \Kcal^2$, $k \neq l$, correspond to the
``trans-dimensional'' part.

The mixture representation~\eqref{equ:transdim-mixture} satisfy the
assumptions of Proposition~\ref{prop:ratio} with $\MM = \Kcal^2$ ,
$\WW_{k,l} = \Xset_k \times \Xset_l$ for all $(k,l) \in \MM$ and
$\varphi(k,l) = (l,k)$. Therefore, if the current state~$x$ is
in~$\Xset_k$ and the proposed state~$x'$ in~$\Xset_l$, the MHG
ratio~\eqref{eq:acc-prob-mix} reads
\begin{equation}
  \label{equ:ratio-transdim-1}
  r(\xx,\xx') \;=\; \frac{ p_{l,k}(\xx') }{ p_{k,l}(\xx) }
  \; \frac{ \pi(\dxx')\, Q_{l,k}( \xx', \dxx) }{
    \pi(\dxx)\, Q_{k,l}( \xx,\dxx') }\, \cdot
\end{equation}
In most ``tutorial'' papers about the RJ-MCMC method, this expression
is directly written in the special case where Green's dimension
matching argument can be applied (see, e.g., \cite{green:2003:trans},
Sections~2.2 and~2.3). Unfortunately, the dimension matching argument
does not apply directly to the commonly used Birth-or-Death kernels
(see next section) if the mixture
representation~\eqref{equ:transdim-mixture}, which leads
to~\eqref{equ:ratio-transdim-1}, is used.

\section{Birth-or-Death kernels}
\label{sec:bod-exchang}

\subsection{Birth-or-Death kernels on (unsorted) vectors}
\label{sec:sec:bod-unsorted}

Let us consider the situation where a point $\xx \in \Xset$ describes
a set of $k$ objects $s_1, \ldots, s_k \in \Sset$, with $\left( \Sset,
  \nu \right)$ an atomless\footnote{See, e.g., \cite{Fremlin2}. As a
  concrete example, think of $\Sset = \Rset^d$ endowed with its usual
  Borel $\sigma$-algebra and $\nu$ equal to Lebesgue's measure. We
  will use the following property in the proof of
  Proposition~\ref{prop:bod-mhg-ratio}: if $(\Sset,\nu)$ is atomless,
  then the diagonal $\Delta = \{ (s,s) : s\in \Sset \}$ is
  $\nu\otimes\nu$-negligible in $\Sset \times \Sset$.} measure space
and $k \in \Nset$. One possible---and commonly used---way of
representing this is to consider pairs $\left( k, \ss \right)$, where
the objects~$s_i$, $1 \le i \le k$, have been arranged in a vector
$\ss = \left( s_1, \ldots, s_k \right) \in \Sset^k$. The corresponding
space is $\Xset = \cup_{k\ge 0} \Xset_k,\; \Xset_k = \{k\} \times
\Sset^k$, with the convention that $\Sset^0 = \{ \varnothing \}$.

\begin{remark}
  The results that will be presented in this section are easily
  generalized if the model includes additional (fixed-dimensional)
  parameters that are left unchanged by the Birth-or-Death moves (for
  instance the parameters~$\Lambda$ and~$\delta^2$ in a fully Bayes
  version of the model presented in Section~\ref{sec:sinusoid}).
\end{remark}

Birth-or-death kernels are the most natural kind of trans-dimensional
moves in such spaces. Given $k\in \Nset$, $\ss = (s_1, \ldots, s_k) \in
\Sset^k$ and $s^* \in \Sset$, we introduce the notations
\begin{align*}
  \ss_{-i} &\;=\; 
  \left( s_1, \ldots, s_{i-1}, s_{i+1}, \ldots, s_k \right)
  \in \Sset^{k-1},
  \\
  \ss \oplus_i s^* &\;=\;  
  \left( s_1, \ldots, s_{i-1}, s^*, s_i, \ldots, s_k \right)
  \in \Sset^{k+1},
\end{align*}
where $1 \le i \le k$ in the first case and $1 \le i \le k+1$ in the
second case. Starting from $\xx = \left( k, \ss \right)$, a birth move
inserts a new component~$s^* \in \Sset$, generated according to some
proposal distribution $q(s)\, \nu(\diffs)$, at a \emph{randomly}
selected location:
\begin{equation}
  \label{equ:b-ker}
  Q_{\bb}\left( \xx, \dotvar \right) \;=\;
  \frac{1}{k+1} \sum_{i=1}^{k+1} \int_\Sset \delta_{(k+1,\ss\oplus_i
    s^*)}\, q(s^*)\, \nu(\diffs^*) \,.
\end{equation}
A death move, on the contrary, removes a \emph{randomly} selected
component form the current state:
\begin{equation}
  \label{equ:d-ker}
  Q_{\dd}\left( \xx, \dotvar \right) \;=\;
  \frac{1}{k} \sum_{i=1}^k \delta_{( k-1, \ss_{-i} )} \,.
\end{equation}
Finally, the birth-or-death kernel is a mixture of the two:
\begin{equation}
  \label{equ:bod-ker}
  Q( \xx, \dotvar ) \;=\; 
  p_{\bb}(\xx)\, Q_{\bb}( \xx, \dotvar ) 
  \,+\,
  p_{\dd}(\xx)\, Q_{\dd}( \xx, \dotvar )\,,
\end{equation}
with $p_{\bb}(\xx) \ge 0$, $p_{\dd}(\xx) \ge 0$, $p_{\bb}(\xx) +
p_{\dd}(\xx) = 1$, and $p_{\dd}\left( (0,\varnothing) \right) =
0$. 

\subsection{Expression of the MHG ratio}
\label{sec:bod-mhg-ratio}

The following proposition provides the expression of the MHG ratio for
the model and kernel described in Section~\ref{sec:sec:bod-unsorted}.

\begin{prop}
  \label{prop:bod-mhg-ratio}
  Assume that, for all~$k \ge 1$, the target measure~$\pi$ restricted
  to~$\Xset_k$ admits a probability density function~$f_k$ with
  respect to~$\nu^{\otimes k}$. Then the MHG ratio is
  \begin{equation}
    \label{eq:bod-mhg-1}
    r(\xx,\xx') \;=\;
    \frac{ f_{k+1}(\xx') }{ f_k(\xx) }
    \;
    \frac{ p_{\dd}(\xx') }{ p_{\bb}(\xx) }
    \;
    \frac{ 1 }{ q(s^*) }
  \end{equation}
  for a birth move from $\xx=(k,\ss)$ to~$\xx' = (k+1,\ss \oplus_i
  s^*)$.
\end{prop}

\medbreak

\begin{proof}  
  Although a direct computation of the MHG ratio would be possible
  based on Equations~\eqref{equ:b-ker}--\eqref{equ:bod-ker}, we find
  it much more illuminating to deduce the result from
  Proposition~\ref{prop:ratio} using kernels which are simpler
  than~$Q_{\bb}$ and~$Q_{\dd}$. To do so, let us consider the family
  of elementary kernels~$Q_m$, with $m$ in the index set
  \begin{equation*}
    \MM = \left\{ 
      (\alpha,k,i) \in \{0,1\} \times \Nset^2:
      1 \le i \le k + \alpha
    \right\}
  \end{equation*}
  where $Q_{1,k,i}$ is the kernel from~$\Xset_k$ to~$\Xset_{k+1}$ that
  inserts a new component $s^* \sim q(s) \nu(ds)$ in position~$i$, and
  $Q_{0,k,i}$ is the kernel from~$\Xset_k$ to~$\Xset_{k-1}$ that
  removes the $i^{\text{th}}$ component. Then we can write
  \begin{equation}
    \label{equ:bod-mixt}
    Q( \xx, \dotvar ) 
    \;=\; 
    \sum_{m\in \MM} j(\xx,m)\, Q_m(\xx,\dotvar),
  \end{equation}
  with $j(\xx,m)$ defined for all $\xx = (k, \ss) \in \Xset$ as 
  \begin{equation*}
    j(\xx,m) \;=\;  \begin{cases}
      p_{\bb}(\xx) / (k+1) & \text{if }
      m=(1,k,i), 1 \le i \le k+1,\\
      p_{\dd}(\xx) / k & \text{if }
      m=(0,k,i), 1 \le i \le k,\\
      0 & \text{otherwise.}
    \end{cases}
  \end{equation*}
  Denote by $\widetilde \Xset_k$ the set of all $\xx \in \Xset_k$ in
  which no two components are equal. For all~$k$, $\pi(\Xset_k
  \setminus \widetilde \Xset_k) = 0$, since $\pi_{|\Xset_k}$ admits a
  density with respect to the product measure~$\nu^{\otimes k}$. The
  mixture representation~\eqref{equ:bod-mixt} thus satisfies the
  assumptions of Proposition~\ref{prop:ratio} with
  \begin{align*}
    \WW_{(1,k,i)} \;=\; \Bigl\{\Bigr.\; & 
    (\xx,\xx') \in \widetilde \Xset_k \times \widetilde \Xset_{k+1}  :\;
    \exists \ss \in \Sset^k,\, \exists s^* \in
    \Sset,\,  \\
    & \quad
    \xx = (k,\ss),\, 
    \xx' = (k+1,\ss\oplus_i s^*) \; 
    \Bigl.\Bigr\},
  \end{align*}
  $\WW_{(0,k,i)} = \WW_{(1,k-1,i)}^\tr$, $\varphi(1,k,i) = (0,k+1,i)$
  and $\varphi(0,k,i) = (1,k-1,i)$. According to
  Proposition~\ref{prop:ratio}, the MHG ratio for a birth move $m =
  (1,k,i)$ is thus
\begin{equation*}
  r(\xx,\xx') 
  \;=\; 
  \frac{ p_{\dd}(\xx') }{ p_{\bb}(\xx) }
  \;
  \frac{%
    \pi(\dxx')\, Q_{0,k+1,i}(\xx', \dxx)
  }{%
    \pi(\dxx)\, Q_{1,k,i}(\xx, \dxx')
  }.
\end{equation*}
Observe that the $1/(k+1)$ terms, in the move selection probabilities,
cancel each other. To complete the proof, it remains to show that
\begin{equation}
  \label{equ:MHG-elem-bod}
  \frac{%
    \pi(\dxx')\, Q_{0,k+1,i}(\xx', \dxx)
  }{%
    \pi(\dxx)\, Q_{1,k,i}(\xx, \dxx')
  }
  \;=\;
  \frac{ f_{k+1}(\xx') }{ f_k(\xx) }
  \;
  \frac{ 1 }{ q(s^*) }\cdot
\end{equation}
This can be obtained, in the general case%
\footnote{In the important special case where $\Sset \subset
  \Rset^d$ and $\nu$ is (the restriction of) the $d$-dimensional
  Lebesgue measure, \eqref{equ:MHG-elem-bod} can be simply seen as the
  result of Green's dimension matching argument
  \cite[Section~3.3]{green:1995:reversible}, in a very simple case
  where the Jacobian is equal to one.}, by a direct computation of
the densities with respect to the symmetric measure
\begin{align*}
  \xi\left (\ddiff(k,\ss), \dxx' \right) \;=\; \nu^{\otimes
    k}(\diffss) \biggl[\; \biggr.
  &    \delta_{(k-1,\ss_{-i})}(\dxx') \\
  & \,+\, \int_{\Sset} \delta_{(k+1,\ss \oplus_i s^*)}\, \nu(\diffs^*)
  \biggl. \;\biggr].
\end{align*}
\end{proof}

  We emphasize that~\eqref{equ:bod-mixt} is \emph{not} the usual
  mixture representation of trans-dimensional kernels introduced in
  Section~\ref{sec:canonical-transdim}. Indeed, starting, e.g.,
  from~$\Xset_k$, there are several elementary kernels that can
  propose a point in~$\Xset_{k+1}$. This shows the usefulness of
  Proposition~\ref{prop:ratio}, which provides sufficient conditions
  for~\eqref{eq:acc-prob-mix} to hold beyond the case of the usual
  mixture representation~\eqref{equ:transdim-mixture}.

\subsection{Birth-or-Death kernels on sorted vectors}
\label{sec:other-repr}

Let us assume now that the objects are ``sorted'', in some sense,
before being arranged in the vector $\ss = \left( s_1, \ldots, s_k
\right) \in \Sset^k$. This happens, in practice, either when there is
a natural ordering on the set of objects (e.g., the jump times in
signal segmentation or multiple change-point problems
\cite{green:1995:reversible, punskaya:2002:bcf}) or when artificial
constraints are introduced to restore identifiability in the case of
exchangeable components (see \cite{richardson:1997:bayesian,
  richardson:1998:corrigendum, cappe:2003:reversible,
  jasra:2005:label, stephens:2000:label} for the case of mixture
models).

To formalize this, let us consider the same space~$\Xset$ as in
Section~\ref{sec:sec:bod-unsorted}. Assume that~$\Sset$ is endowed
with a total order and that the corresponding ``sort function''
$\psi:\Xset \to \Xset$ is measurable. What we are assuming now is that
the target measure, denoted by~$\widetilde{\pi}$ in this section, is
supported by~$\psi(\Xset)$---in other words, the components of~$\xx
\in \Xset$ are $\widetilde{\pi}$-almost surely sorted.

In such a setting, the definition of the Birth-or-Death kernel has to
be slightly modified in order to accommodate the sort constraint: the
death kernel is unchanged, but new components are inserted
\emph{deterministically} at the only location that makes the resulting
vector sorted (instead of being added at a random
location). Mathematically, for $\xx = (k,\ss) \in \Xset_k$, we now
have:
\begin{align*}
  \widetilde{Q}_{\bb}\left( \xx, \dotvar \right) 
  & \;=\;
  \int_\Sset \delta_{\psi(k+1,\ss\oplus_1
    s^*)}\, q(s^*)\, \nu(\diffs^*) \,,\\
  \widetilde{Q}_{\dd}\left( \xx, \dotvar \right)
  & \;=\;
  \frac{1}{k} \sum_{i=1}^k \delta_{( k-1, \ss_{-i} )} 
  \;=\; Q_{\dd}\left( \xx, \dotvar \right) \,.  
\end{align*}
Proceeding as in the proof of Proposition~\ref{prop:bod-mhg-ratio}, it
can be proved that the MHG ratio for a birth move from $\xx=(k,\ss)$
to~$\xx' = (k+1,\ss \oplus_i s^*)$ is
\begin{equation}
  \label{eq:mhg-sorted}
    r(\xx,\xx') \;=\;
    \frac{ \widetilde{f}_{k+1}(\xx') }{ \widetilde{f}_k(\xx) }
    \,\cdot\,
    \frac{ p_{\dd}(\xx') / (k+1) }{ p_{\bb}(\xx)\, \eta_{i}(\xx) }
    \,\cdot\,
    \frac{ 1 }{ q(s^*)/\eta_{i}(\xx) } \,,
\end{equation}
where $\widetilde{f}_k$ denotes the pdf of~$\widetilde{\pi}$
on~$\Xset_k$ and $\eta_{i}(\xx)$ the probability that $s^* \sim q(s)\,
\nu(\diffs)$ is inserted at location~$i$ in~$\xx$. (Note that
$p_{\bb}(\xx)\, \eta_{i}(\xx)$ is the probability of performing a
birth move at location~$i$, and $p_{\dd}(\xx') / (k+1)$ the
probability of the reverse death move; this is the appropriate way of
decomposing this kernel as mixture in order to use
Proposition~\ref{prop:ratio}.)

Let us now consider the case where, in the setting of
Section~\ref{sec:sec:bod-unsorted}, the target probability
measure~$\pi$ is invariant under permutations of the components
indices (in other words, the corresponding random variables are
\emph{exchangeable} \cite[Chapter~4]{bernardo2000bayesian}). Sorting
the components (as an identifiability device) is equivalent to looking
at the image measure~$\widetilde{\pi} = \pi^{\psi}$, which has the pdf
$\widetilde{f}_k \;=\; k!\, f_k\, \un_{\psi(\Xset)}$ on $\Xset_k$. As
a consequence, the MHG ratios~\eqref{eq:bod-mhg-1}
and~\eqref{eq:mhg-sorted} are equal.

\begin{remark}
  Another option, when the components of the vector $\left( s_1,
    \ldots, s_k \right)$ are exchangeable, is to forget about the
  indices and consider the set~$\left\{ s_1, \ldots, s_k \right\}$
  instead. The object of interest is then a (random) finite set of
  points in~$\Sset$---in other words, a point process on~$\Sset$. The
  expression of the MHG ratio for Birth-or-Death moves in the point
  process framework, with the Poisson point process as a reference
  measure, has been given in~\cite{geyer94spp} (one year before the
  publication of Green's paper \cite{green:1995:reversible}). Point
  processes have been widely used, since then, in image processing and
  object identification (see, e.g., \cite{rue:1999:bayesian,
    descombes:2004:object, stoica:2004:gibbs,lacoste:2005:point}).
\end{remark}

\section{Example: joint detection and estimation of sinusoids
  in white Gaussian noise}
\label{sec:sinusoid}

The results presented in Section~\ref{sec:bod-exchang} can be used to
compute the MHG ratio easily in many signal decomposition
problems. Let us illustrate this with the joint Bayesian model
selection and parameter estimation of sinusoids in white Gaussian
noise, as first considered by~\cite{andrieu:1999:jbm}. As explained in
the introduction, this seminal paper introduced the RJ-MCMC
methodology in the signal processing community, and at the same time
introduced an erroneous expression of the MHG ratio that has been,
since then, reproduced in a long series of papers. We follow closely
the model and notations of~\cite{andrieu:1999:jbm}; the reader is
referred to the original paper for more details.

Let $\y \,{=}\, \left( y_{1},\, y_{2},\, \ldots,\, y_{N} \right)^t$ be
a vector of $N$ observations of an observed signal. We consider the
finite family of nested models $\model_0 \subset \model_1 \subset
\cdots \subset \model_{\kmax}$, where $\model_k$ assumes that $\y$ is
composed of~$k$ sinusoids observed in white
Gaussian noise. Let $\ok = \left( \omega_{1,k}, \ldots, \omega_{k,k}
\right)$ and $\ak = \left( a_{c_{1,k}}, a_{s_{1,k}}, \ldots,
  a_{c_{k,k}}, a_{s_{k,k}} \right)$ be the vectors of radial
frequencies and cosine/sine amplitudes under model~$\model_k$,
respectively; moreover, let $\D_k$ be the corresponding $N \times 2k$
design matrix.  Then, the observed signal~$\y$ follows
under~$\model_k$ a normal linear regression model:
\begin{equation*}   
  \y = \vect{D}_k .\ak + \vect{n},
\end{equation*}
where $\vect{n}$ is a white Gaussian noise with variance $\sigma^2$.
The unknown parameters are, then, assumed to be the number of
components $k$, the component-specific parameters $\thk = \{ \ak, \ok
\}$ and the noise variance $\sigma^2$ which is common to all
models. The joint prior distribution is chosen to have the following
hierarchical structure:
\begin{equation*}
  p\left( k, \thk, \sigma^2 \right) 
  \;=\; 
  p\bigl( \ak \mid k, \ok, \sigma^2 \bigr)\;
  p\bigl( \ok \mid k \bigr) \;
  p\bigl( k \bigr)\; 
  p\bigl( \sigma^2 \bigr),
\end{equation*}
where the prior over $\ak$ is the conventional $g$-prior
distribution~\cite{zellner:1986:g-prior}, which is a zero mean
Gaussian with $\sigma^2 \ds\; (\D_k^t \D_k)^{-1}$ as its covariance
matrix. Conditional on~$k$, the radial frequencies are independent and
identically distributed, with a uniform distribution on $(0,
\pi)$. The noise variance~$\sigma^2$ is endowed with Jeffreys improper
prior, i.e. {$p(\sigma^2) \propto 1 / \sigma^2$}. The number of
components $k$ is given a Poisson distribution with mean~$\Lambda$,
truncated to~$\left\{ 0, 1, \ldots, \kmax \right\}$. The
parameters~$\ak$ and~$\sigma^2$ can be integrated out analytically,
and the resulting marginal posterior becomes
\begin{equation}
  \label{equ:target}
  p\left( k, \ok \,{\mid}\, \y \right)
  \,\propto\; \, (\yPy{k})^{-N/2}
  \frac{\Lambda^k \pi^{-k}}{k!\, (\ds+1)^k}\,
  \un_{(0,\pi)^k}(\ok)\,,
\end{equation}
with
\vspace{-8pt}
\begin{equation*}
  \vect{P}_k \;=\; \vect{I}_N \,-\,
  \frac{\ds}{1+\ds}\, \vect{D}_k 
  \left( \DtD \right)^{-1} \vect{D}_k^t
\end{equation*}  
when $k \geq 1$ and $\vect{P}_0 = \vect{I}_N$.  

Inference under this hierarchical Bayesian model is carried out
in~\cite{andrieu:1999:jbm} using an RJ-MCMC sampler on~$\Xset =
\bigcup_{k=0}^{\kmax}\, \{k\} \times \left( 0 ,\, \pi \right)^k$ with
target density~\eqref{equ:target}. We only focus here on the
``between-models'' moves, which are Birth-or-Death moves of the kind
described in Section~\ref{sec:sec:bod-unsorted}, with a uniform
density on~$(0,\,\pi)$ for the proposal distribution of the new
frequency in the birth moves.

Let us now compute the MHG ratio for a birth move. Note that the
posterior density~\eqref{equ:target} is written in the case of
``unsorted'' components described in
Sections~\ref{sec:sec:bod-unsorted}--\ref{sec:bod-mhg-ratio}. We shall
therefore make use of Proposition~\ref{prop:bod-mhg-ratio}, which
assumes that new component is inserted at a random position~$i$ (all
components being selected with the same probability). The correct MHG
ratio, for a birth move from~$\xx = (k,\, \ok)$ to~$\xx' = (k+1, \,
\ok\oplus_i\omega^*)$, turns out to be
\begin{equation}
  \label{eq:birth-ratio-sin-0}
  r(\xx, \xx') \;=\;%
  \frac{ p\left( k+1, \ok\oplus_i\omega^* \,{\mid}\, \y \right)}%
  { %
    p\left( k, \ok \,{\mid}\, \y \right)} \times
  \frac{p_{\dd}(\xx')}{p_{\bb}(\xx)} \times
  \frac{1}{q\left(\omega^*\right)},
\end{equation}
where $q$ denotes the uniform distribution of~$\left( 0, \pi
\right)$. Using
\begin{equation*}
  \frac{p_{\dd}(\xx')}{p_{\bb}(\xx)}
  \;=\; \frac{ p_0(k) }{ p_0(k+1) }
  \:=\; \frac{ k+1 }{ \Lambda }
\end{equation*}
as in \cite{andrieu:1999:jbm}, with $p_0$ standing for the (truncated
Poisson) prior distribution of~$k$, we finally find
\begin{align}
  \label{eq:birth-ratio-sin}
  r(\xx, \xx') \;=\; %
  & \left(\frac{\yPy{k+1}}{\yPy{k}}\right)^{-N/2} \; \frac{\Lambda
    \pi^{-1}}{(1+k)(1+\ds)}
  \nonumber \\%
  & \;\times\; \frac{k+1}{\Lambda} %
  \;\times\; \frac{1}{ \pi^{-1} } %
  \nonumber \\%
  \;=\; &
  \left(\frac{\yPy{k+1}}{\yPy{k}}\right)^{-N/2} \; \frac{1}{1+\ds}
  \,\cdot
\end{align}
Note that the expression of the ratio proposed in \cite[Equation
(20)]{andrieu:1999:jbm} differs from the one we find here by a factor
$1 / (k+1)$. A similar mistake in computing RJ-MCMC ratios has been
reported in the field of genetics~\cite{jannink04mh,
  sillanpaa:2004:comment}.

\begin{figure}
  \centering \input{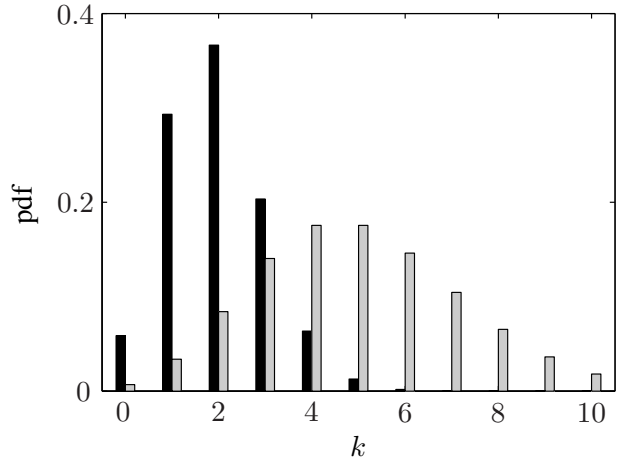}
  \caption{The pdf's of Poisson (gray) and accelerated Poisson
    (black) distributions with mean~$\Lambda \,=\, 5$. Both
    distributions are truncated to the set $\{0, \ldots, 32\}$.}
  \label{fig:poiss}
\end{figure}

In fact, using the expression of the birth ratio with an additional
factor of $1 / (k+1)$, as in~\cite{andrieu:1999:jbm}, amounts to
assigning a different prior distribution over~$k$ called ``accelerated
Poisson distribution''~\cite{sillanpaa:2004:comment} which reads
\begin{equation}\label{eq:acc_poiss}
  p_2(k) \;\propto\; \frac{e^{-\Lambda}\Lambda^k}{(k!)^2} \,
  \un_\Nset(k) .
\end{equation}
Figure~\ref{fig:poiss} illustrates the difference between both the
accelerated (black) and the usual (gray) Poisson distributions when
mean~$\Lambda \,=\, 5$. It can be observed that the accelerated
Poisson distribution~\eqref{eq:acc_poiss} puts a stronger emphasis on
``sparse'' models, i.e., models with a small number of components.

\begin{figure}
  \centering \input{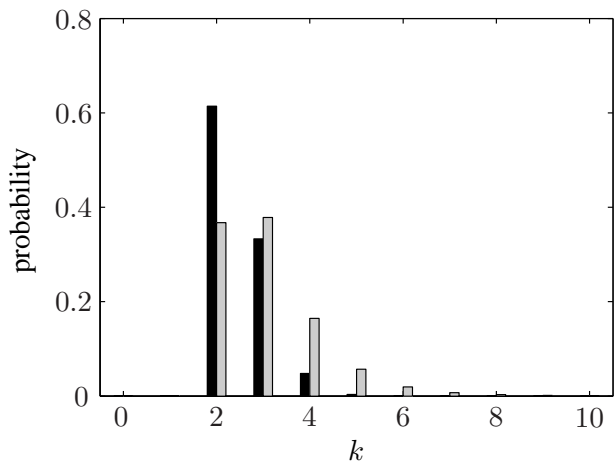}
  \caption{Frequency of selection for each model~$\model_k$ for 100
    replications of the experiment described in
    Section~\ref{sec:sinusoid}, using the expression of the ratio
    given in~\cite[Equation (20)]{andrieu:1999:jbm} (black) and the
    corrected ratio~\eqref{eq:birth-ratio-sin} (gray). There are~$k=3$
    sinusoidal components in the observed signal~$\y$ and
    the~$\text{SNR} = 7 \text{dB}$. 100k samples were generated using
    RJ-MCMC sampler and the first 20k were discarded as burn-in
    period.}
  \label{fig:ratio}
\end{figure}

Let us consider an experiment in which the observed signal of
length~$N=64$ consists of~$k = 3$ sinusoidal components with the
radial frequencies~$\ok\,=\,(0.63, 0.68, 0.73)^t$ and
amplitudes~$a_{c_{i,k}}^2 + a_{s_{i,k}}^2 = (20, 6.32, 20)^t$, $1\leq
i \leq k$. The signal to noise ratio, defined as $\text{SNR}
\triangleq \|\D_k . \ak\|^2\, /\, (N \sigma^2)$, is set to a
moderate value of $7\text{dB}$. Samples from the posterior
distribution of~$k$ are obtained using the RJ-MCMC sampler
of~\cite{andrieu:1999:jbm}, with an inverse Gamma
prior~$\mathcal{IG}(2,100)$ on~$\delta^2$ and a Gamma
prior~$\mathcal{G}(1,10^{-3})$ on~$\Lambda$. For each observed signal
in 100 replications of the experiment, the sampler was run twice: once
with the correct expression of the ratio, given
by~\eqref{eq:birth-ratio-sin}, and once with the erroneous expression
from \cite{andrieu:1999:jbm}. Figure~\ref{fig:ratio} shows the
frequency of selection of each model under both the Poisson and the
accelerated Poisson distribution as a prior for~$k$. It appears that
the (unintended) use of the accelerated Poisson distribution, induced
by the erroneous expression of the MHG ratio, can result in a
significant shift to the left of the posterior distribution of~$k$.

\begin{remark}
  Working with ``sorted'' vectors of frequencies would be quite
  natural in this problem, since the frequencies are exchangeable
  under the posterior~\eqref{equ:target}. As explained in
  Section~\ref{sec:other-repr}, the expression of the MHG ratio would
  be the same.
\end{remark}

\begin{remark}
  The reason why the MHG ratio in \cite{andrieu:1999:jbm} is wrong can be
  understood from a subsequent paper \cite{andrieu:2001:msm}, where
  the same computation is explained in greater detail. There we can
  see that the authors, working with an ``unsorted vector''
  representation, consider that the new component in a birth move is
  \emph{inserted at the end}. The death move, however, is defined as
  in the present paper: a sinusoid to be removed is \emph{selected
    randomly} among the existing components. Here is the mistake: if
  the new component is inserted at the end during a birth move, then
  any attempt at removing a component which is not the last one should
  be rejected during a death move. In other words, the acceptance
  probability should be zero when any component but the last one is
  picked to be removed during a death move.
\end{remark}

\section{Conclusion}
\label{sec:conclusion}

The computation of MHG ratios is a delicate matter involving
measure-theoretic considerations, for which practitioners need clear
mathematical statements that can be used ``out of the box''. Such a
statement has been available for a long time in the classical
fixed-dimensional Metropolis-Hastings sampler, and more recently
provided by Green \cite{green:1995:reversible} for trans-dimensional
moves that comply with the assumptions of his dimension matching
argument. 

In this note, we have provided the expression of the MHG ratio for
Birth-or-Death moves, using a general result for mixtures of proposal
kernels, and corrected the erroneous expression provided by
\cite{andrieu:1999:jbm}. A similar correction has to be applied to the
ratios used in the long series of signal processing papers
\cite{andrieu:2000:RJMCMC-SA, andrieu:2001:msm, andrieu:2001:robust,
  andrieu:2002:bayesian, larocque:2002:reversiblejump,
  larocque:2002:particle, ng:2005:wideband, davy:2006:bap,
  shi:2007:study, melie:2008:bayesian, ng:2008:particle,
  hong:2010:joint, schmidt:2010:inmf, rubtsov:2007:time} that have
been found to contain the same mistake.

While writing this note, we discovered that a very similar mistake had
been detected and corrected in the field of genetics by
\cite{jannink04mh}, from which we borrow our concluding words:
\emph{The fact that this error has remained in the literature for over
  5~years} [12~years in the present case] \emph{underscores the view
  that while Bayesian analysis using Markov chain Monte Carlo is
  incredibly flexible and therefore powerful, the devil is in the
  details. Furthermore, incorrect analyses can give results that seem
  quite reasonable.}

\bibliographystyle{plainnat}
\bibliography{refs}

\end{document}

%% file: figures/poiss.tex
%
%
\begin{psfrags}%
\psfragscanon%
%
\psfrag{s01}[b][b]{\color[rgb]{0,0,0}\setlength{\tabcolsep}{0pt}\begin{tabular}{c}pdf\end{tabular}}%
\psfrag{s02}[t][t]{\color[rgb]{0,0,0}\setlength{\tabcolsep}{0pt}\begin{tabular}{c}$k$\end{tabular}}%
%
\psfrag{x01}[t][t]{$0$}%
\psfrag{x02}[t][t]{$2$}%
\psfrag{x03}[t][t]{$4$}%
\psfrag{x04}[t][t]{$6$}%
\psfrag{x05}[t][t]{$8$}%
\psfrag{x06}[t][t]{$10$}%
%
\psfrag{v01}[r][r]{$0$}%
\psfrag{v02}[r][r]{$0.2$}%
\psfrag{v03}[r][r]{$0.4$}%
%
\includegraphics[width=8cm]{poiss.eps}%
\end{psfrags}%
%

%% file: figures/ratio7dB.tex
%
%
\begin{psfrags}%
\psfragscanon%
%
\psfrag{s01}[b][b]{\color[rgb]{0,0,0}\setlength{\tabcolsep}{0pt}\begin{tabular}{c}probability\end{tabular}}%
\psfrag{s02}[t][t]{\color[rgb]{0,0,0}\setlength{\tabcolsep}{0pt}\begin{tabular}{c}$k$\end{tabular}}%
%
\psfrag{x01}[t][t]{$0$}%
\psfrag{x02}[t][t]{$2$}%
\psfrag{x03}[t][t]{$4$}%
\psfrag{x04}[t][t]{$6$}%
\psfrag{x05}[t][t]{$8$}%
\psfrag{x06}[t][t]{$10$}%
%
\psfrag{v01}[r][r]{$0$}%
\psfrag{v02}[r][r]{$0.2$}%
\psfrag{v03}[r][r]{$0.4$}%
\psfrag{v04}[r][r]{$0.6$}%
\psfrag{v05}[r][r]{$0.8$}%
%
\includegraphics[width=8cm]{ratio7dB.eps}%
\end{psfrags}%
%